\documentclass{ws-procs975x65}
\newcommand{\be}{\begin{equation}}
\newcommand{\ee}{\end{equation}}
\newcommand{\bea}{\begin{eqnarray}}
\newcommand{\eea}{\end{eqnarray}}
\usepackage[spanish,english]{babel}
\usepackage[latin1]{inputenc}
\usepackage{epsfig}

\begin{document}

\title{{\bf Avoiding the Big Bang Singularity with Palatini $f(R)$ Theories. }}

\author{Carlos Barragán}
\address{Departamento de Física Teórica, Universidad Autónoma de Madrid,  28049 Madrid, Spain}

\author{Gonzalo J. Olmo}
\address{Instituto de Estructura de la Materia, CSIC, Serrano 121, 28006 Madrid, Spain}

\author{Hèlios Sanchis-Alepuz}
\address{ Fachbereich Theoretische Physik, Institut für Physik, Karl-Franzens-Universität Graz,
Universitätsplatz 5, A-8010 Graz, Austria}

\begin{abstract}
We show that there exist modified theories of gravity in which the metric satisfies second-order equations and in which the Big Bang singularity is replaced by a cosmic bounce without violating any energy condition. In fact, the bounce 
is possible even for presureless dust. We give a  characterization of such 
theories, which are formulated in the Palatini formalism, and discuss their dynamics in the region near the bounce.  We consider spatially flat and non-flat homogeneous and isotropic universes. 
\end{abstract}

\bodymatter
\vspace{1cm}
The Einstein equivalence principle supports the idea that gravitation is a curved spacetime phenomenon. However, this principle, grounded on extremely solid experimental facts, does not point uniquely towards general relativity (GR) or any other theory of gravity as the {\it most reasonable or sensible} theory of gravity. It took a deep and thorough understanding of the classical laws of physics and the genius of Einstein to combine this principle and a set of second order differential equations for the spacetime metric to construct GR, one of the most successful theories of all times. 
A common characteristic of most of the theories proposed as alternatives  to GR is the existence in them of new degrees of freedom. Some theories contain new dynamical fields (scalars, vectors, or tensors of different ranks), others introduce higher order derivatives of the metric field, and others represent combinations of these two possibilities. Such theories are usually proposed to change some aspect of GR at a particular regime, though the result is usually more involved. The introduction of new degrees of freedom (via new fields or via higher-order equations), leads to the emergence of new solutions which may depart from those of GR not only locally (at the particular regime where the modification was supposed to act) but also globally, i.e., in other regimes where the GR solution should hold. An example of this undesired behavior is found in $f(R)$ theories in metric formalism, where infrared corrections not only affect the late-time cosmology but may also lead to important changes in the solar system\cite{Olmo07a} or in earlier phases of the cosmic history\cite{Amendola:2006we}.\\
\indent If an $f(R)$ theory is formulated in the Palatini formalism, in which the connection is independent of the metric, then the resulting field equations for the metric are still second-order, like in GR. These theories lead to modified gravitational dynamics not because of the emergence of new dynamical degrees of freedom but rather because of the enhanced gravitational effects of matter induced by the independent connection, which turns out to be fully determined by the metric and the matter. Because of the lack of extra degrees of freedom, the resulting dynamics is only modified at the scale chosen, being all other regimes almost unaffected, though some models may have dramatic effects\cite{Olmo08a}. The solutions are thus very close to those of GR but with certain deformations at specific regimes, which depend on the parameters of the Lagrangian $f(R)$. In this talk, we summarize some important results found recently that show that Palatini $f(R)$ theories with high curvature corrections are able to turn the big bang singularity of GR into a big bounce, thus regularizing the GR solutions in this particular regime. In the case of homogeneous and isotropic universes, we also show that the conditions for a bounce to exist are quite insensitive to the  existence of non-zero spatial curvature\footnote{This corrects an error made in our previous work\cite{BOSA09} regarding universes with non-zero spatial curvature}.\\        
\indent The action that defines Palatini $f(R)$ theories is of the form 
$S=\frac{1}{2\kappa^2}\int d^4x\sqrt{-{g}}f({R})+S_m[{g}_{\mu \nu},\psi_m]$. 
Here $f({R})$ is a function of ${R}\equiv{g}^{\mu \nu }R_{\mu \nu }(\Gamma )$, with $R_{\mu \nu }(\Gamma )$ given by
$R_{\mu\nu}(\Gamma )=-\partial_{\mu}
\Gamma^{\lambda}_{\lambda\nu}+\partial_{\lambda}
\Gamma^{\lambda}_{\mu\nu}+\Gamma^{\lambda}_{\mu\rho}\Gamma^{\rho}_{\nu\lambda}-\Gamma^{\lambda}_{\nu\rho}\Gamma^{\rho}_{\mu\lambda}$
where $\Gamma^\lambda _{\mu \nu }$ is an independent connection (not the Levi-Civita connection of $g_{\mu \nu }$). Assuming a general homogeneous and isotropic Universe, the relevant Friedmann equations are\cite{BOSA09}
\begin{eqnarray}\label{eq:Hubble-iso-f(R)}
H^2&=&\frac{1}{6f_R}\frac{\left[f+(1+3w)\kappa^2\rho-\frac{6K f_R}{a^2}\right]}{\left[1+\frac{3}{2}\tilde\Delta_1\right]^2} \\
\label{eq:expansion-f(R)}
[2+3\tilde\Delta_1]\dot H&+&3\left[2+(2-3w)\tilde\Delta_1+3\tilde\Delta_2\right]H^2=\frac{\left[f+2\kappa^2P-\frac{4K f_R}{a^2}\right]}{f_R} \ , 
\end{eqnarray}
where $\tilde{\Delta}_1=-(1+w)\rho\partial_\rho f_R/f_R=(1+w)(1-3w)\kappa^2\rho  f_{RR}/(f_R(Rf_{RR}-f_R))$, $\tilde{\Delta}_2=(1+w)^2\rho^2\partial_{\rho\rho} f_R/f_R $, $f_R\equiv \partial_R f$, and $R=R(\rho)$ follows from the relation $Rf_R-2f=-\kappa^2(1-3w)\rho$. To derive these equations we assumed a perfect fluid with constant equation of state $P=w\rho$.\\
The conditions for a cosmic bounce follow from $H^2=0$. In the models considered so far in the literature\cite{Olmo-Singh09,BOSA09}, this happens typically when $f_R$ vanishes at some value of $R$. The vanishing of $f_R$ implies a divergence of $\tilde{\Delta}_1\sim 1/f_R$, which pushes $H^2$ to zero as $\sim f_R\to 0$. 
In the region near the bounce, we find that (\ref{eq:expansion-f(R)}) can be approximated by $\dot H+(2-3w)H^2=R_B/[6(1-3w)]$, where $R_B$ is the curvature at the bounce. Defining, like in GR, $R_B=(1-3w)\kappa^2\rho_B$, we see that the right hand side of this equation is positive, which guarantees that $H$ has reached a minimum. The solutions to these equation are straigthforward ($\alpha^2\equiv \kappa^2\rho_B/6$):
\begin{equation}
H^2(t)_{w<\frac{2}{3}}= \frac{\alpha^2 \tanh^2[\alpha\sqrt{2-3w}(t-t_B)]}{2-3w}  \ , \ H^2(t)_{w>\frac{2}{3}}= \frac{\alpha^2\tan^2[\alpha\sqrt{3w-2}(t-t_B)] }{3w-2} \nonumber
\end{equation}
where $t_B$ is an integration constant that sets the instant at which the bounce occurs. The corresponding expansion factor is given by
\begin{equation}
a(t)_{w<\frac{2}{3}}= a_B (\cosh[\alpha\sqrt{2-3w}(t-t_B)])^\frac{1}{2-3w}  \ , \
a(t)_{w>\frac{2}{3}}= a_B (\cos[\alpha\sqrt{3w-2}(t-t_B)])^\frac{1}{2-3w} \nonumber
\end{equation}
These results represent the generic behavior for any Lagrangian $f(R)$ near the bounce caused by the condition $f_R=0$ irrespective of the value of $K$.\\
\indent Besides bounces due to $f_R=0$, other possibilities also seem possible, such as theories with $(Rf_{RR}-f_R)\to 0$ at some $R\equiv R_B$. In this case one finds $\dot{H}=[(2R_B)/(9(1+w))](f_{RRR}/f_{RR})|_{R_B}$, which is constant and independent of $K$. Unfortunately, no satisfactory example with that behavior is yet known. In fact, though the model $f(R)=R_P(e^{R/R_P}-1)$ leads to  $(Rf_{RR}-f_R)\equiv e^{R/R_P} (R/R_P-1)$, with a zero at $R=R_P$, the denominator of $H^2$ develops singularities before reaching that point if $w<1/3$. For $w>1/3$ a bounce arises as $R/R_P\approx\ln(1-\rho/\rho_P) \to-\infty$ due to the vanishing of $f_R=e^{R/R_P}$. A third possibility arises in universes with $K>0$ which corresponds to the vanishing of $\left[f+(1+3w)\kappa^2\rho-\frac{6K f_R}{a^2}\right]$ in (\ref{eq:Hubble-iso-f(R)}).\\
\indent {Acknowledgments.} H.S-A. has been partially supported by the Austrian Science
Fund FWF under Project No. P20592-N16. G.J.O. thanks MICINN for a Juan de la Cierva contract, the Spanish Ministry of Education and its program ``José Castillejo'' for funding a stay at the CGC of the University of Wisconsin-Milwaukee. G.J.O. has also been partially supported by grant FIS2008-06078-C03-02. 

\begin{figure}[t]
\begin{center}
\psfig{file=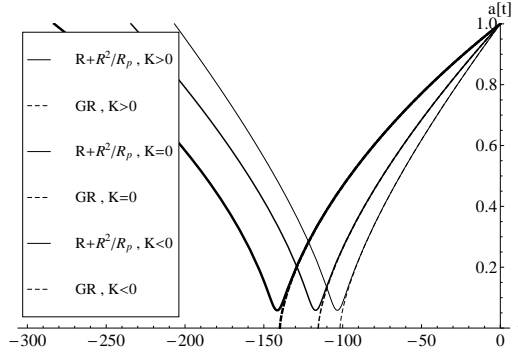,width=0.6\textwidth}
\end{center}
\caption{Bouncing solutions for the model $f(R)=R+R^2/R_P$ compared with the GR solutions for $w=0$. The three curves represent $K>0,K=0,$ and $K<0$ solutions (from left to right) respectively. The $K>0$ solution yields cyclic bounces both in the future and in the past.}
\label{aba:fig1}
\end{figure}

\end{document}